\documentclass[12pt]{article}

\usepackage{epsfig,latexsym,amsfonts,amsmath,amsthm,amssymb,amsbsy,multirow,slashed,wasysym,textcomp,subfigure,hyperref,wrapfig,datetime,comment,mathtools,cancel,cite}
\usepackage[font={footnotesize,sl},bf]{caption}

\def\ee{\end{equation}}
\def\bea{\begin{eqnarray}}
\def\eea{\end{eqnarray}}
\newcommand{\beq}{\begin{eqnarray}}
\newcommand{\eqq}{\end{eqnarray}}
 \newcommand{\badat}{\begin{alignedat}}
 \newcommand{\eadat}{\end{alignedat}}

\newcommand{\eal}[1]{\be \begin{aligned} #1 \end{aligned}\end{equation}} 
\newcommand{\eqn}[1]{\be #1 \end{equation}} 
\newcommand{\eqa}[1]{\bea  #1\end{eqnarray}}

\long\def\new#1\endnew{{\bf #1}}		
\long\def\del#1\enddel{}

\def\del{\partial}

\usepackage{color}

\newcommand{\pink}[1]{\textcolor{\pink}{#1}}

\definecolor{dblue}{rgb}{0.2,0.50,0.80} 

 \newcommand{\virg}{\hspace{1 mm}, \hspace{8 mm}}
\newcommand{\p}{\partial}

\def\O{\mathcal{O}}

\def\bh{{\bar h}}
\def\bz{{\bar z}}
\def\bw{{\bar w}}

\def\scri{{\mathcal{I}}}

\numberwithin{equation}{section} % equation numbers follow sections

\begin{document}

 \begin{titlepage}
  \thispagestyle{empty}
  \begin{flushright}
  CPHT-RR097.092018
  \end{flushright}
  \bigskip
  \begin{center}
	 \vskip2cm
  \baselineskip=13pt {\LARGE \scshape{Conformally Soft Photons and Gravitons}}
  % \vskip1cm \today
	 \vskip1cm
   \centerline{
   {Laura Donnay}${}^\diamondsuit{}^\odot$ 
   {, Andrea Puhm}${}^\blacklozenge{}^\diamondsuit{}^\odot$
   {and Andrew Strominger}{}${}^\diamondsuit{}^\odot$
   }
	
 \bigskip
  %\vskip1cm
  \end{center}

\begin{abstract}
  \noindent
	
  The four-dimensional $S$-matrix  is reconsidered as a correlator on the celestial sphere at null infinity. Asymptotic particle states can  be characterized by the point at which they enter or exit the celestial sphere as well as their $SL(2,\mathbb C)$ Lorentz quantum numbers: namely their conformal scaling dimension and spin $h\pm \bar h$ instead of  the energy and momentum.  This characterization precludes the notion of a  soft particle whose energy is taken to zero. We propose it should be replaced by the notion of a {\it conformally soft}  particle with $h=0$ or $\bar h=0$. For photons we explicitly construct conformally soft $SL(2,\mathbb C)$ currents with dimensions $(1,0)$ and identify them with the generator of a $U(1)$  Kac-Moody symmetry  on the celestial sphere. For gravity the generator of celestial conformal symmetry is constructed from a $(2,0)$ $SL(2,\mathbb C)$ primary wavefunction. 
 Interestingly, BMS supertranslations are generated by  a spin-one weight  $(\frac{3}{2},\frac{1}{2})$ operator, which nevertheless shares holomorphic characteristics of  a conformally soft operator. This is  because the right hand side of its OPE with a weight $(h,\bar h)$   operator ${\cal O}_{h,\bar h}$ involves the shifted operator ${\cal O}_{h+\frac{1}{2},\bar h+ \frac{1}{2}}$. This OPE relation looks quite unusual from the celestial CFT$_2$ perspective but is equivalent to the leading soft graviton theorem and may usefully constrain celestial correlators in quantum gravity. 
\end{abstract}

%\vskip3cm
\vfill
\noindent{\em${}^\diamondsuit$ Center for the Fundamental Laws of Nature, Harvard University,}
{\em Cambridge, MA 02138, USA}

\noindent{\em${}^\odot$ Black Hole Initiative, Harvard University,}
 {\em Cambridge, MA 02138, USA}
 
\noindent{\em${}^\blacklozenge$       
      Centre de physique th\'{e}orique, \'{E}cole Polytechnique, CNRS,}
{\em 91128, Palaiseau, France}

 \end{titlepage}
%\tableofcontents
\section{Introduction}

In the conventional formulation of quantum field theory (QFT) an important role is played by soft particles whose energy $\omega\to0$.  Scatttering amplitudes containing such particles obey special relations which are central to the consistency of  QFT\cite{Yennie:1961ad,Weinberg:1965nx}. Recently, a reformulation of 4D QFT has been explored in which asymptotic particle states are described by $SL(2,\mathbb C)$-Lorentz primary wavefunctions instead of the usual energy-momentum eigenstates
\cite{deBoer:2003vf,Strominger:2013lka,He:2015zea,Pasterski:2016qvg,Cheung:2016iub,Strominger:2017zoo,Pasterski:2017kqt,Pasterski:2017ylz,Schreiber:2017jsr,Lam:2017ofc,Banerjee:2017jeg,Stieberger:2018edy}. Such wavefunctions are labelled by their $SL(2,\mathbb C)$ conformal dimensions $(h,\bar h)$ (related to the twist) and their asymptotic positions on the celestial sphere, while the $S$-matrix takes the form of a correlation function in a celestial 2D CFT. %Such a reformulation of QFT scattering is of particular interest for a potential holographic 
%formulation of 4D quantum gravity in MInkowski space, but may have other applications as well. 
In general one expects that some features of scattering will be easier to understand in the new formulation while others will become harder. 

One thing which is lost in the new formulation is the notion of a soft particle. $SL(2,\mathbb C)$ primary wavefunctions are not energy eigenstates so the energy cannot be taken to zero. Instead, we have the notion of a {\it conformally soft} particle for which the conformal dimension either $h$ or $\bar h$ is taken to zero. The symmetries of the celestial sphere imply that the scattering of such particles also obey special relations. 

In this paper we will construct several interesting examples of conformally soft particles. 
In an elegant recent paper Pasterski and Shao \cite{Pasterski:2017kqt}  showed for photons that the conformal primary wavefunctions in unitary principal series with dimensions $(1+\frac{i\lambda}{2},\frac{i\lambda}{2})$ form a complete basis (for one helicity), and that the $\lambda\to 0$ wavefunction is the Goldstone mode for spontaneously broken large gauge symmetries. This is an example of a conformally soft particle which simply decouples from all scattering amplitudes. However the canonical partner of the Goldstone mode was not considered in the discussion of  \cite{Pasterski:2017kqt}. Here we show that an additional logarithmic branch of the solution space appears for $\lambda \to 0$, and construct from it the missing canonical partner of the Goldstone mode. This conformally soft mode does not decouple from scattering amplitudes. Rather it generates a Kac-Moody symmetry on the celestial sphere \cite{Strominger:2013lka,He:2015zea} which can be identified with the large gauge symmetry of QED. 

The situation is even more interesting for gravitons. For this case it was shown \cite{Pasterski:2017kqt}  that the conformal primary wavefunctions in the unitary principal series with dimensions $(\frac{3}{2}+\frac{i\lambda}{2},-\frac{1}{2}+\frac{i\lambda}{2})$ form a complete basis (for one helicity). Taking $\lambda\to 0$, we again get a Goldstone mode for spontaneously broken BMS supertranslation symmetry
which decouples from scattering. Again a logarithmic branch appears with a canonically conjugate wavefunctions which has dimension
$(\frac{3}{2},-\frac{1}{2})$ and  enters into the soft part of the supertranslation charge. This is not conformally soft in the sense stated above since both left and right dimensions are nonzero. However we see that a suitable divergence of this wavefunction is related to the dimension $(\frac{3}{2},\frac{1}{2})$ current $P_z$ \cite{Strominger:2013jfa,Barnich:2013axa} which generates supertranslations on the celestial sphere. More specifically, OPEs with $P_z$ take one operator to a second canonically related one with dimensions increased by $(\frac{1}{2},\frac{1}{2})$. Hence OPEs involving $P_z$ have the $1\over z-w$ factor characteristic of a holomorphic current.  Finally we consider wavefunctions with $\lambda=-i$ and hence dimensions $(2,0)$. A suitable convolution of these wavefunctions with the field operator at null infinity gives precisely the known formula \cite{Kapec:2016jld} for the 2D stress tensor on the celestial sphere, previously obtained by reverse engineering from the subleading soft graviton theorem \cite{Cachazo:2014fwa}. 

The paper is organized as follows. We start in Section~\ref{sec:MinkSphere} by introducing a map between Minkowski space and the celestial sphere at null infinity. In Section~\ref{sec:softphoton}, we construct the conformally soft photon, which has conformal dimensions $(1,0)$, and show that it is the canonical partner of the Goldstone mode associated to large gauge transformations at null infinity. In Section~\ref{sec:softgraviton}, we turn to the gravity case and construct a $(\frac{3}{2},-\frac{1}{2})$ wavefunction and show that it is the canonical partner of the spin-two Goldstone mode, and is related to the current which generates supertranslations. We finally construct a conformally soft graviton mode of conformal dimensions $(2,0)$ and discuss its relation to the 2D stress tensor for 4D gravity. In the Appendix, we collect details about the inner product of conformal primaries, the shadow transform, and provide explicit expressions for the conformally soft and Goldstone modes at null infinity.

\section{Minkowski $\rightarrow$ Celestial Sphere}\label{sec:MinkSphere}
Let $X^\mu$, with $\mu=0,1,2,3$, be the Cartesian coordinates\footnote{Our signature convention is $\eta_{\mu \nu}=\rm{diag}(-,+,+,+)$.} on the four-dimensional Minkowski spacetime $\mathbb R^{1,3}$. Massless particles exit flat spacetime at future null infinity where they intersect the asymptotic  sphere at infinity. This sphere is referred to as the celestial sphere and denoted $\mathcal{CS}^2$.
A natural map between Minkowski spacetime and the celestial sphere is obtained by going to Bondi coordinates $(u,r,z,\bz)$:
\begin{equation}
 X^0=u+r\,, \quad X^i=r \hat{X}^i(z,\bz)\,, \quad \hat{X}^i(z,\bz)=\frac{1}{1+z \bar z}(z+\bz,-i(z-\bz),1-z\bz)\,.
\end{equation}
In these coordinates the Minkowski line element is
\begin{equation}\label{ret}
ds^2=-du^2-2du dr+2r^2 \gamma_{z \bar z} dz d\bar z\,,
\end{equation}
where $u$ is the retarded time, $r$ is the radial coordinate and $z$ is a complex coordinate on the unit sphere with metric
\begin{equation}
\gamma_{z \bar z}=\frac{2}{(1+z \bar z)^2}\,.
\end{equation}

A massless particle crosses the celestial sphere at a point $(w,\bw)$ with momentum \mbox{$p^\mu=\frac{\omega}{1+w\bw}q^\mu(w,\bw)$}, with $q^\mu(w,\bw)$ a null vector and $\omega \geq0$ the energy. The null vector $q^\mu$ as a function of $w,\bar w$ is
\begin{equation}\label{qmu}
 q^{\mu}(w,\bw)=(1+w\bw,w+\bw,-i(w-\bw),1-w\bw)\,.
\end{equation} 
Under an $SL(2,\mathbb C)$ transformation 
\begin{equation}
w \to \frac{a w+b}{cw+d} \,, \quad  \bar w \to \frac{ \bar a  \bar w+ \bar b}{ \bar c \bar w+ \bar d}\,,
\label{SL2C}
\end{equation} 
where $ad-bc=1$, $q^\mu$ transforms as a vector up to a conformal weight,
\begin{equation}
 q^{\mu}\to  q^{\mu'} =(cw+d)^{-1}(\bar c\bar w+\bar d)^{-1}\Lambda^\mu_{\,\,\, \nu}q^\nu\,,
\end{equation} 
 and $\Lambda_\mu^{\,\;\nu}$ is the associated $SL(2,\mathbb{C})$ group element in the four-dimensional representation\footnote{For an explicit expression for $\Lambda_\mu^{\,\;\nu}$ in terms of $a,b,c,d$ see for instance~\cite{Oblak:2015qia}.}. 
Note that null vectors \eqref{qmu} satisfy
\begin{equation}
 q^{\mu}(w,\bw) q_{\mu}(w',\bw')=-2|w-w'|^2\,,
\end{equation} 
and the derivative of~\eqref{qmu} with respect to $w$ ($\bw$) is the photon polarization vector $\epsilon^\mu_w$ ($\epsilon^\mu_\bw$) of positive (negative) helicity:
\begin{equation}
\partial_w q^\mu=\sqrt 2 \epsilon^\mu_w(q)=(\bw,1,-i,-\bw) \quad \,, \quad \partial_\bw q^\mu=\sqrt 2 \epsilon^\mu_\bw(q)=(w,1,i,-w)\,,
\end{equation}
satisfying 
\begin{equation}
 \epsilon_w\cdot q=0\,, \quad \epsilon_w\cdot \epsilon_w=0\,, \quad \epsilon_w\cdot \epsilon_\bw=1\,,
\end{equation}
and similarly for $w\leftrightarrow \bw$. The completeness relationship is
\begin{equation}
\epsilon_{w}^\mu \epsilon_{\bw}^\nu+\epsilon_w^\mu \epsilon_{\bw}^\nu=\eta^{\mu \nu} +\frac{1}{2}(q^\mu n^\nu+n^\mu q^\nu)\,,
\end{equation}
with
$n^\mu=\partial_w \partial_\bw q^\mu=(1,0,0,-1)$.  The graviton polarization tensor of positive (negative) helicity is $\epsilon^{\mu\nu}_{ww}=\epsilon^\mu_w \epsilon^\nu_w$ ($\epsilon^{\mu\nu}_{\bw\bw}=\epsilon^\mu_\bw \epsilon^\nu_\bw$).  

\section{Conformally Soft Photons}\label{sec:softphoton}

\subsection{Massless Spin-One Conformal Primary}
The outgoing $(+)$ and incoming $(-)$ massless spin-one conformal primary wavefunctions (\mbox{$X^\mu \in \mathbb{R}^{1,3}$} and $a=w,\bw$ is the index on the celestial sphere) are\footnote{The $i \varepsilon$-prescription is added to circumvent the singularity at $q\cdot X=0$.}~\cite{Cheung:2016iub,Pasterski:2017kqt} 
\begin{eqnarray}\label{ADelta}
\badat{2}
A_{\mu;a}^{\Delta,\pm}(X^\mu;w,\bw)&=\frac{\partial_a q_\mu}{(-q \cdot X \mp i \varepsilon)^{\Delta}}+\frac{(\partial_a q\cdot X) \, q_\mu}{(-q \cdot X \mp i \varepsilon)^{\Delta+1}}\,,
\eadat
\end{eqnarray}
where $q^\mu$ is a function of $(w,\bar w)$ as given in \eqref{qmu}.
They transform as two-dimensional conformal primaries with conformal dimensions $(h,\bar h)=\frac{1}{2}(\Delta+J,\Delta-J)$ 
under an $SL(2,\mathbb{C})$ Lorentz transformation:
\begin{equation}
A_{\mu;a}^{\Delta,\pm}\left(\Lambda^\rho_{\,\, \nu}X^\nu;\frac{a w+b}{c w+d},\frac{\bar a \bw+\bar b}{\bar c \bw+\bar d}\right)=(c w+ d)^{2h}(\bar c \bw+ \bar d)^{2\bh}\Lambda_\mu^{\,\; \sigma}A_{\sigma;a}^{\Delta,\pm}(X^\rho;w,\bw)\,.
\label{SL2Cspin1}
\end{equation}
When the index  $a=w$ the  spin $J=+1$ (positive helicity) while for $a=\bw$,  $J=-1$ (negative helicity).
The spin-one conformal primary wavefunctions satisfy both the radial and the Lorenz gauge conditions 
\begin{equation}
 X^\mu A_{\mu;a}^{\Delta,\pm}=0\,, \quad \partial^\mu  A_{\mu;a}^{\Delta,\pm}=0\,,
\label{gaugecond}
\end{equation}
and are solutions to the four-dimensional Maxwell equations
\begin{equation}
\partial_\rho \partial^\rho A_{\mu;a}^{\Delta,\pm}=0\,.
\end{equation}
It is convenient to decompose, following \cite{Pasterski:2017kqt,Pasterski:2017ylz}, \eqref{ADelta} as 
\begin{eqnarray}
\badat{2}
A_{\mu;a}^{\Delta,\pm}&=\frac{\Delta-1}{\Delta(\mp i)^\Delta \Gamma(\Delta)}\, {V}_{\mu;a}^{\Delta,\pm}  +\partial_\mu \alpha_a^{\Delta,\pm}\,,
\label{prim1}
\eadat
\end{eqnarray}
where
\begin{equation}\label{hatValpha}
{V}_{\mu;a}^{\Delta,\pm} (X^\mu;w,\bw)  =(\mp i)^\Delta \Gamma(\Delta)\frac{\partial_a q_\mu}{(-q \cdot X \mp i \varepsilon)^\Delta}\,, \quad \alpha_a^{\Delta,\pm} (X^\mu;w,\bw) = \frac{\partial_a q \cdot X }{\Delta (-q \cdot X \mp i \varepsilon)^\Delta}\,.
\end{equation}
The residual gauge transformation $\alpha^{\Delta,\pm}_a$ preserves the Lorenz gauge condition $\partial^2 \alpha^{\Delta,\pm}_a=0$. 
${V}^{\Delta,\pm}_{\mu;a}$ satisfies the Lorenz gauge condition but not the radial gauge condition. It is gauge equivalent to the conformal primary wavefunction \eqref{prim1} when $\Delta \neq 1$ and, with the given normalization, is the Mellin transform of the canonically normalized plane wave\footnote{The Mellin transform of a function $f(\omega)$ is defined by $\bar{f}(\Delta)=\int_0^\infty d\omega \omega^{\Delta-1} f(\omega)$ and its inverse transformation is $f(\omega)=\int_{-\infty}^\infty \frac{d\Delta}{2\pi i} \omega^{-\Delta} \bar{f}(\Delta)$ with $\omega>0$. The $i\varepsilon$ prescription in~\eqref{VMellinSpin1} is necessary to regulate the integral. Unless it is important for the discussion we will omit the regulator.}
\begin{equation}\label{VMellinSpin1}
{V}_{\mu;a}^{\Delta,\pm}(X^\mu;w,\bw)=
\partial_a q_\mu \int_0^\infty d\omega \, \omega^{\Delta-1} e^{\pm i\omega q\cdot X-  \varepsilon \omega}\,.
\end{equation}
Notice that ${V}^{\Delta}$ does not transform covariantly under $SL(2,\mathbb{C})$ but the non-covariant terms are pure residual gauge, hence, following~\cite{Pasterski:2017kqt}, we will still call them conformal primaries.

\subsubsection{Shadow Transform}

The shadow transform, which involves a convolution in $w$,  maps a primary wavefunction with conformal dimension~$\Delta$ to a primary wavefunction with conformal dimension $2-\Delta$. The shadow transform of the spin-one conformal primary wavefunction $A_{\mu;a}^{\Delta}$~\eqref{ADelta} was computed in~ in terms of its uplift $A_{\mu;\nu}^\Delta$ to the embedding space $\mathbb{R}^{1,3}$; we review the definition of the shadow transform and this computation in Appendix~\ref{app:shadow}. The result for the shadow wavefunction is \cite{Pasterski:2017kqt}\footnote{In order to define the transform  along the light-cone $-X^2=0$ and $q\cdot X=0$ we need to prescribe a proper regulator. This can be achieved by an imaginary timelike shift of $X^\mu \to X^\mu \pm i \varepsilon V^\mu$ with $V^\mu=(-1,0,0,0)$. We are grateful to S. Pasterski for discussions on this point.}.
\begin{equation}
%   \widetilde{A^{\Delta}_{\mu;a^*}}=(-X^2)^{1-\Delta} A_{\mu;a}^{2-\Delta}\equiv \widetilde{A}^{2-\Delta}_{\mu;a}\,.
\widetilde{A}^{2-\Delta,\pm}_{\mu;a}=(-X^2)^{1-\Delta} A_{\mu;a}^{2-\Delta,\pm}\,.
 \label{ShADelta}
\end{equation}
One can verify that the shadow transform~\eqref{ShADelta} of a spin one conformal primary is itself a spin-one conformal primary.

\subsubsection{Conformal Basis}

A conserved inner product between complex spin-one wavefunctions is 
\begin{equation}\label{IPSpin1}
(A,A')=-i\int d^3 X^i [A^j F'^{*}_{0j}-A'^{j*}F_{0j}]\,,
\end{equation}
where $i=1,2,3$ is the spatial index in $\mathbb R^{1,3}$ and $*$ denotes the complex conjugation.
It was shown by Pasterski and Shao in~\cite{Pasterski:2017kqt} that, with respect to the inner product~\eqref{IPSpin1}, the conformal primary wavefunctions ${V}^{\Delta}$ form a $\delta$-function-normalizable basis on the principal continuous series $\Delta=1+i \lambda$ with $\lambda \in \mathbb R$:
\begin{eqnarray}\label{VVSpin1}
\badat{2}
({V}_{\mu;a}^{\Delta,\pm}(X^\mu;w,\bw),{V}_{\mu;a'}^{\Delta',\pm}(X^\mu;w',\bw'))
&=\pm (2\pi)^4 \,\delta(\lambda-\lambda') \delta_{aa'}\,\delta^{(2)}(w-w')\,,
\eadat
\end{eqnarray}
up to zero-mode issues which we will discuss below. Alternatively, an equally good basis of spin-one conformal primary wavefunctions is spanned by the shadow transformed conformal primaries $\widetilde V^\Delta$.
The right-hand side of~\eqref{VVSpin1} is obtained by taking Mellin transforms of the inner product of plane waves; we review the computation in Appendix~\ref{app:IP}.
With a similar computation one finds that the inner product for the spin-one conformal primary wavefunctions $A^{\Delta}$ has an additional normalization factor: 
\begin{eqnarray}\label{AA}
\badat{2}
(A_{\mu;a}^{\Delta,\pm}(X^\mu;w,\bw),A_{\mu;a'}^{\Delta',\pm}(X^\mu;w',\bw'))
&=\frac{\lambda \sinh(\pi \lambda)e^{\mp \pi \lambda}}{\pi(1+\lambda^2)} ({V}_{\mu;a}^{\Delta,\pm}(X^\mu;w,\bw),{V}_{\mu;a'}^{\Delta',\pm}(X^\mu;w',\bw'))\,.
\eadat
\end{eqnarray}

We now turn to the important subtleties arising at $\lambda=0$. The canonically normalized $V^\Delta$ inner product~\eqref{VVSpin1} pairs $\Delta=1+i\lambda$ modes with their $\Delta=1-i\lambda$ partners (recall that the definition~\eqref{IPSpin1} involves a complex conjugation). However, for $\Delta=1$, the primaries $V^\Delta$ are ill-defined and so is the inner product~\eqref{VVSpin1}. 
To obtain a complete basis of conformal primary wavefunctions on the principal continuous series $\Delta=1+i \lambda$ with $\lambda \in \mathbb R$, we need to include a canonically paired set of zero modes. One of them is readily obtained by the $\Delta\to1$ limit of the conformal primary wavefunction $A^\Delta$.
In this limit, the mode $A^\Delta$ becomes a pure gauge transformation which  can be recognized as the antipodally matched large gauge transformation of \cite{He:2014cra}. This identifies $A^{\Delta=1}$ as the Goldstone mode associated to the spontaneously broken large gauge symmetries. However its canonical partner appears to be missing. 

In a momentum space decomposition, the canonical partner of the Goldstone mode is known to be  the $\omega \to 0$ soft photon \cite{He:2014cra}. 
We will show in the next section that the missing partner of the Goldstone mode can be identified from a subtle logarithmic branch of the solution space arising at $\Delta=1$. 

\subsection{$\Delta=1$ Conformal Modes}\label{sec:csphoton}

\subsubsection{Goldstone Mode}
In the limit $\Delta \to 1$, the conformal primary wavefunction $A_{\mu;a}^{\Delta}$ \eqref{ADelta} and its shadow $\widetilde{A}_{\mu;a}^{2-\Delta}$ \eqref{ShADelta} coincide and reduce to a total derivative. 
We identify this pure gauge spin-one conformal primary with the Goldstone mode:
\begin{equation}\label{goldstone}
A^{\rm G}_{\mu;a}\equiv \lim_{\Delta \to 1}A^{\Delta,\pm }_{\mu;a}= \partial_\mu \alpha_{a}^{1}\,,
 \end{equation}
where 
\begin{equation}
 \alpha_{a}^{1} =-\frac{\partial_a q \cdot X}{q \cdot X}\,,
 \label{alpha1}
\end{equation}
is the $\Delta\to 1$ limit of the pure gauge parameter~\eqref{hatValpha}, and does not depend on the $\pm i\varepsilon$  prescription. 

\subsubsection{Conformally Soft Mode}
The Goldstone mode at $\Delta=1$ is so far missing a  canonical partner. Fortuitously, solutions to Maxwell's equations that are not pure gauge can be constructed from the following combination of $A^\Delta$ and its shadow:\footnote{The $\pm$ superscript refers to $i \varepsilon$ regulators at $-X^2=0$ and $q\cdot X=0$ given explicitly in the fully regulated logarithmic mode.}
\begin{equation}\label{logmode}
A^{{\rm log ,\pm}}_{\mu;a} \equiv \lim_{\Delta \to 1}\partial_\Delta \left(A^{\Delta,\pm}_{\mu;a}+\tilde{A}^{2-\Delta,\pm}_{\mu;a}\right)  \,. 
\end{equation}
The conformal transformation of the mode~\eqref{logmode} is that of a $\Delta=1$ primary wavefunction
\begin{equation}
A_{\mu;a}^{\rm log,\pm}\to (c w+ d)^{1+J}(\bar c \bw+\bar d)^{1-J}\Lambda_\mu^{\,\; \nu}A_{\nu; a}^{{\rm log ,\pm}}.
\end{equation}
The fully regulated logarithmic mode is
\begin{equation}
A_{\mu;a}^{\rm log,\pm}=-{\rm log} \left[-X^2\mp 2 i \varepsilon X^0-\varepsilon^2\right]\p_\mu\left(\frac{\p_a (q \cdot X\pm i\varepsilon q^0)}{-q\cdot X \mp i \varepsilon q^0} \right).
\end{equation}
The presence of the logarithm is natural: when two linearly independent solutions to a differential equation degenerate a logarithmic solution typically appears (see for instance~\cite{Grumiller:2008qz}).
Its field strength $F^{{\rm log},\pm}_{\mu\nu;a}=\partial_\mu A^{{\rm log},\pm}_{\nu;a}-\partial_\nu A^{{\rm log},\pm}_{\mu;a}$ is
\begin{equation}
F^{{\rm log},\pm}_{\mu\nu;a}=-\frac{2(X_\mu \pm i \varepsilon \delta_\mu^0)}{X^2\pm2i \varepsilon X^0+\varepsilon^2}\, \p_\nu\left(\frac{\p_a (q \cdot X\pm i\varepsilon q^0)}{-q\cdot X \mp i \varepsilon q^0} \right)-(\mu \leftrightarrow \nu)\,.
\label{FlogregR2}
\end{equation}
We are interested in the dimension $(1,0)$ difference of these two log modes which we call the conformally soft (CS) photon,
\begin{equation}
 F^{\rm CS}_{\mu\nu;a}\equiv \frac{1}{2\pi i}\left( F^{{\rm log},+}_{\mu\nu;a}-F^{{\rm log},-}_{\mu\nu;a}\right)\,.
\end{equation}
This will be shown below to form a canonical pair with the Goldstone mode \eqref{goldstone}. (The sum of the two log modes decouples from the Goldstone mode and is ill-behaved at infinity. Henceforth it is  ignored.)
For regions in spacetime in which either $X^2=0$ or $q \cdot X=0$, but not both at the same time, we use the representation of the delta function and its derivative for $\varepsilon \to 0$
\begin{equation}
\badat{2}
 &\delta(x)=-\frac{1}{2\pi i} \left(\frac{1}{x+i \varepsilon}-\frac{1}{x-i \varepsilon}\right)\,,\\
& \delta'(x)= \frac{1}{2\pi i}\left(\frac{1}{(x+i \varepsilon)^2}-\frac{1}{(x-i \varepsilon)^2}\right)= -\frac{\delta(x)}{x},
\eadat
\end{equation}
and obtain a distributional expression when $\varepsilon \to 0$ for the conformally soft photon:
\begin{equation}
F^{\rm CS}_{\mu\nu;a}=2 X_\mu A^{\rm G}_{\nu;a} \left(\delta\left(X^2\right)+\frac{(q\cdot X)}{X^2} \delta(q\cdot X)\right)-(\mu \leftrightarrow \nu).
\label{FCS}
\end{equation}
One can directly verify that $\partial^\mu F^{\rm CS}_{\mu\nu;a}=0$. 
The conformally soft gauge field, such that $ F^{\rm CS}_{\mu\nu;a}=\p_\mu A^{\rm CS}_{\nu;a}-(\mu \leftrightarrow \nu)$, is given by 
\begin{equation}\label{ACS}
 A^{\rm CS}_{\mu;a}=(q \cdot X) \log[X^2]A^{\rm G}_{\mu;a} \delta(q\cdot X) +A^{\rm G}_{\mu;a}\Theta\left(X^2\right),
\end{equation}
and transforms as a $\Delta=1$ conformal primary wavefunction.

\begin{figure}[ht!]
\begin{center}
 \includegraphics[width=0.4\textwidth]{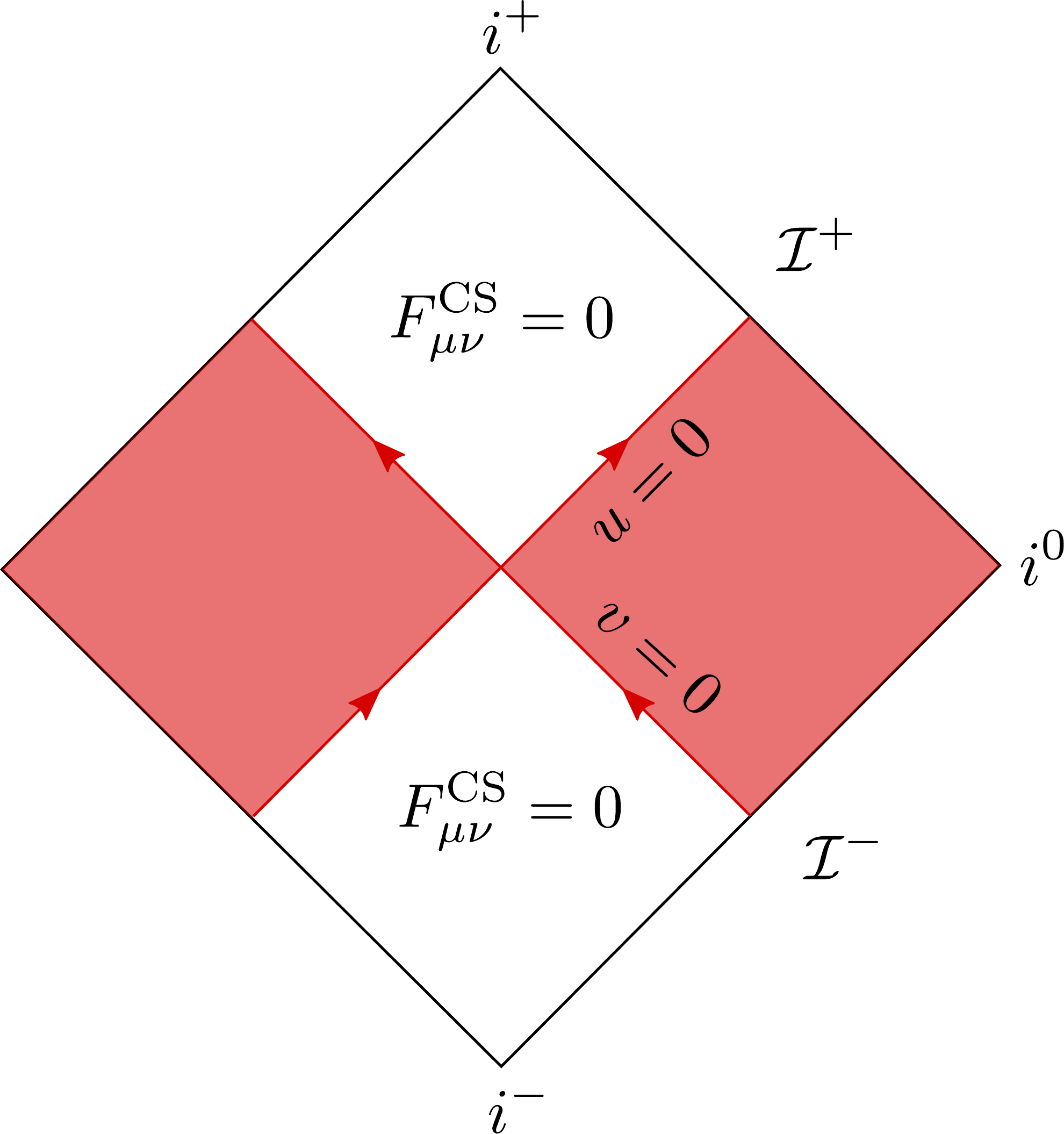} 
 \end{center}
 \caption{Wavefunction of the conformally soft photon. 
 A radiative shock wave with initial data at $v=0$ emerges from past null infinity $\scri^-$, impinges on the origin, and reemerges at $u=0$ at future null infinity  $\scri^+$. Coulombic fields are produced by and confined to the future of the incoming shockwave, and annihilated by and confined to the past of the outgoing shock wave.}
 \label{shock}
\end{figure}

The solution~\eqref{FCS} represents a radiative shock wave which comes in along the past light cone of the origin and emerges along the future light cone. In the intervening regions  Coulomb fields appear at the locus of $q\cdot X=0$, which lies outside (or on) the light cone of the origin.
Inside the past or future light cone all fields vanish. It is illuminating to look at the behavior near null infinity, denoted by $\scri$, and hence use the retarded coordinates $(u,r,z,\bz)$ in which the line element for Minkowski spacetime is given by~\eqref{ret}. 
%$\scri^+$ is the null surface obtained by taking the limit $r\to\infty$ while keeping $(u,z,\bz)$ fixed. We use the symbol $\scri^+_+$ ($\scri^+_-$) to denote the future (past) boundary of $\scri^+$ at $(u=\infty,z,\bz)$ ($(u=-\infty,z,\bz)$). 
Near $\scri$, fields can be expanded in powers of $1/r$; one finds (details are given in Appendix \ref{app:Bondi}) that the expansion of the $z$ component of the Goldstone mode $A^{\rm G}_{z;w}=\partial_z \alpha^1_w$, at both future null infinity $\scri^+$ and past null infinity $\scri^-$, is given by
\begin{equation}\label{scrigoldstone}
A^{\rm G}_{z;w}=-\frac{1}{(z-w)^2}\,, %\alpha^{1}=\frac{1}{z-w}+\mathcal O(1/r)\,,
 \end{equation}
while the $\bz$ component of the Goldstone mode $A^{\rm G}_{\bz;w}=\partial_\bz \alpha^1_w$ develops a $\delta$-function:
\begin{equation}
A^{\rm G}_{\bz;w}=2\pi \delta^{(2)} (z-w)\,.
\end{equation}
The field strength of the Goldstone mode of course vanishes. 
The $uz$ and $u\bz$ components of the conformally soft mode field strength at leading order at $\scri^+$ are
\begin{equation}\label{out}
 F^{{\rm CS}}_{uz;w}= \frac{\delta(u)}{(z-w)^2}\,, \quad F_{u\bz;w}^{{\rm CS}}=-2\pi \delta(u)\delta^{(2)}(z-w)\,,
\end{equation}
while at $\scri^-$ (with the usual antipodal identification of the celestial coordinates)
\begin{equation}
 F^{{\rm CS}}_{vz;w}= \frac{\delta(v)}{(z-w)^2}\,, \quad F_{v\bz;w}^{{\rm CS}}=-2\pi \delta(v)\delta^{(2)}(z-w)\,.
\end{equation}
This last expression is the incoming initial data for a radiative shock wave which impinges on the origin and then, according to \eqref{out}
reemerges at $u=0$. 
These field strengths do not on their own satisfy the constraint equation and Bianchi identity 
\begin{equation}
\badat{2}
 &r^2 \partial_u F_{ur}-\gamma^{z\bar z}(\partial_\bz F_{uz}+\partial_z F_{u\bz})=0,\\
&\partial_u F_{z\bz}+\partial_\bz F_{uz}-\partial_z F_{u\bz}=0\,,
 \eadat
\end{equation}
which require the following expressions for Coulombic fields on $\scri^+$
\begin{equation} 
\badat{2}\label{urzzb}
 r^2 F^{\rm CS}_{ur;w}= 4\pi \gamma^{z\bar z}\partial_z \delta^{(2)}(z-w) \Theta(-u)\,, \quad F^{\rm CS}_{z\bz;w}=0\,.
 \eadat
\end{equation}
Similarly, to satisfy
\begin{equation}
\badat{2}
 &r^2 \partial_v F_{vr}+\gamma^{z\bar z}(\partial_\bz F_{vz}+\partial_z F_{v\bz})=0,\\
&\partial_v F_{z\bz}+\partial_\bz F_{vz}-\partial_z F_{v\bz}=0\,,
 \eadat
\end{equation}
on $\scri^-$ requires the Coulombic fields
\begin{equation}
\badat{2}
 r^2 F^{\rm CS}_{vr;w}= 4\pi \gamma^{z\bar z}\partial_z \delta^{(2)}(z-w) \Theta(v)\,, \quad F^{\rm CS}_{z\bz;w}=0\,.
 \eadat
\end{equation}
These Coulombic fields are produced by and confined to the future of the incoming radiative shock wave along $v=0$, and annihilated by and confined to the past of the outgoing shock wave along $u=0$. 
In summary this describes the wavefunction of the conformally soft photon; see Figure \ref{shock}.

It is interesting to  project the solution~\eqref{FCS} onto its self dual part 
\begin{equation}
 F^{\rm SD}_{\mu \nu}\equiv \frac{1}{2}(F_{\mu \nu}^{\rm CS}-i (^*F^{\rm CS})_{\mu \nu} )\,,
\end{equation}
leading to the following values at $\scri^+$:
\begin{equation}
\badat{2}
 &F^{{\rm SD}}_{uz;w}=\frac{\delta(u)}{(z-w)^2}\,,\quad  F_{u\bz;w}^{{\rm SD}}=0,\\
&  r^2 F^{\rm SD}_{ur;w}=-\gamma^{z\bar z} F^{\rm SD}_{z\bz;w}= 2\pi \gamma^{z\bar z}\partial_z \delta^{(2)}(z-w) \Theta(-u)\,,
 \eadat
\end{equation}
and similarly one can obtain the projection onto its anti-self dual part.

\subsubsection{Canonical Pairing}
We now show that the new conformally soft mode is the canonical partner of the Goldstone mode with respect to the inner product~\eqref{IPSpin1}.
The inner product is independent of the choice of the surface we integrate over, and taking the integral in~\eqref{IPSpin1} to be over $\scri^+$, we find that the inner product between the Goldstone and the conformally soft mode is:
\begin{equation}
i(A^{{\rm CS}}_w(w), A^{\rm G}_{w'}(w'))_{\scri^+}= 8\pi^2 \delta^{(2)}(w-w')\,.
\end{equation}
Moreover, from~\eqref{AA} it follows that the inner product of the conformally soft mode with any other conformal primary $V^{\Delta}$ vanishes. Hence, the Goldstone mode and the conformally soft mode,~\eqref{goldstone} and~\eqref{ACS}, are the pair of zero-modes that enhance the conformal primary wavefunctions $V^\Delta$ to a $\delta$-function-normalizable basis on the principal continuous series $\Delta=1+i \lambda$ with $\lambda \in \mathbb{R}$,  including the point $\lambda=0$.

\subsubsection{Quantum Currents}
 
The inner product enables us to associate a mode of the quantum field operator $\hat A$ to every classical solution of the wave equation. The mode expansion of $\hat A$ on the complete basis of spin-one conformal primary wavefunctions on the principal continuous series $\Delta=1+i\lambda$ with $\lambda\in \mathbb{R}$ is 
\begin{equation}
\badat{2}\label{modeexp}
 \hat{A}_\mu (X) &= \int \frac{d^2 w\, d\lambda \sqrt{2}}{(2\pi)^4 } \,\left({a}_{\lambda w} V^{*1-i\lambda,+}_{\mu;\bw}(X)+ {a}^{\dagger}_{\lambda \bw} V^{1+i\lambda,-}_{\mu;w}(X)+(w \leftrightarrow \bw )\right)\\
  & \quad +\int \frac{d^2 w}{8\pi^2} \,\left(S_w A^{\rm G}_{\mu;w}(X)+ J_w A^{{\rm CS}}_{\mu;w}(X)+(w \leftrightarrow \bw )\right)\,,
 \eadat
\end{equation}
where ${a}_{\lambda w}$ and ${a}^{\dagger}_{\lambda \bw}$ are respectively the annihilation and creation operators of photons obeying the commutation relation
\begin{equation}
 [{a}_{\lambda w}(w),{a}^{\dagger}_{\lambda' w'}(w')]= \frac{1}{2}(2\pi)^4  \delta(\lambda-\lambda')\delta^{(2)}(w-w')\,,
\end{equation}
and similarly, ${J}_w, {S}_w$ obey
\begin{equation}
 [{J}_w(w),{S}_{w'}(w')] = 8\pi^2 \delta^{(2)}(w-w')\,.
\end{equation}
They can be expressed in terms of the inner products of the field operator $\hat{A}$ with respectively $A^{\rm G}$ and $A^{\rm CS}$.
Let us first consider the operator associated to the Goldstone mode
 \begin{equation}
 J_w= i(\hat{A},A^{\rm G}_w).
 \end{equation}
 One may immediately see that quantum commutators with $J_w$ generate large gauge transformations on $\scri^+$:
 \begin{equation} 
 [J_w,\hat A_z]=A_{z;w}^{\rm G}=-{1 \over (z-w)^2}\,.
 \end{equation}
The soft part of the large gauge charge can be expressed as weighted integrals of $J_w$ over the sphere. Combining $J_w$ with its $\scri^-$ counterpart gives  the soft photon current \cite{He:2014cra,Strominger:2017zoo}. It is a dimension $(1,0)$ current whose insertions  generate a $U(1)$ current algebra on the celestial sphere.
We may also consider the operator associated to the conformally soft mode
\begin{equation}
 S_w= i(\hat{A},A^{\rm CS}_w).
\end{equation}
 This is related to (the $\scri^+$ part of) the Goldstone current of \cite{Nande:2017dba}.

\section{Conformally Soft Graviton}\label{sec:softgraviton}
\subsection{Massless Spin-Two Conformal Primary}
The outgoing $(+)$ and incoming $(-)$ massless spin-two conformal primary wavefunctions in $\mathbb{R}^{1,3}$ are~\cite{Pasterski:2017kqt}
\begin{equation}
h_{\mu\nu;a}^{\Delta,\pm}(X^\mu;w,\bw)=\frac{1}{2}\frac{[(-q\cdot X)\partial_{a} q_{\mu}+(\partial_{a} q\cdot X)q_{\mu}][(-q\cdot X)\partial_{a} q_{\nu}+(\partial_{a} q\cdot X)q_{\nu}]}{(-q\cdot X\mp i \varepsilon)^{\Delta+2}}\,.
\label{hDelta}
\end{equation}
Primary wavefunctions \eqref{hDelta} solve the vacuum linearized Einstein equation and are symmetric in the four-dimensional vector indices and symmetric and traceless in the two-dimensional vector indices. They transform as both a four-dimensional rank-two tensor and as two-dimensional spin-two conformal primaries with conformal dimension $(h,\bar h)=\frac{1}{2}(\Delta+J,\Delta-J)$ under an $SL(2,\mathbb{C})$ Lorentz transformation:
\begin{equation}
h_{\mu \nu;a}^{\Delta,\pm}\left(\Lambda^\mu_{\,\, \nu}X^\nu;\frac{a w+b}{c w+d},\frac{\bar a \bw+\bar b}{\bar c \bw+\bar d}\right)=(c w+d)^{\Delta+J}(\bar c \bw +\bar d)^{\Delta-J}\Lambda_{\mu}^{\,\, \rho}\Lambda_{\nu}^{\,\, \sigma}h_{\rho 
\sigma;a}^{\Delta,\pm}(X^\mu;w,\bw)\,.
\label{SL2Cspin2}
\end{equation}
The two-dimensional index $a=ww$ corresponds to spin $J=+2$ (positive helicity) while $a=\bw\bw$ corresponds to $J=-2$ (negative helicity).

The conformal primary wavefunction~\eqref{hDelta} is traceless and satisfies Lorenz and radial gauge conditions:
\begin{equation}
\eta^{\mu \nu}h_{\mu \nu;a}^{\Delta,\pm}=0 \quad \,, \quad \partial^\mu h_{\mu \nu;a}^{\Delta,\pm}=0\quad \,, \quad X^\mu h_{\mu \nu;a}^{\Delta,\pm}=0\,,
\label{gauge}
\end{equation}
and hence the vacuum linearized Einstein equations become
\begin{equation}
\partial^\rho \partial_\rho h_{\mu \nu;a}^{\Delta,\pm}(X^\mu;w,\bw)=0\,.
\end{equation}
A representative wavefunction that differs from~\eqref{hDelta} only by a pure diffeomorphism is \cite{Pasterski:2017kqt}\footnote{Here and hereafter in similar contexts it is understood that the  $a$ index on the rhs is $ww$ or $\bw\bw$, while the pair of $a$s on the rhs are each a single $w$ or $\bw$.}
\begin{equation}
 {V}_{\mu\nu;a}^{\Delta,\pm}(X^\mu;w,\bw)=(\mp i)^\Delta \Gamma(\Delta) \frac{\frac{1}{2}\partial_{a} q_{\mu} \partial_{a} q_{\nu}}{(-q\cdot X\mp i \varepsilon)^\Delta}\,.
 \label{VMellinSpin2}
\end{equation}
It is traceless and satisfies the Lorenz gauge condition but not the radial gauge condition. This representative conformal primary has the advantage of being related to plane waves by a Mellin transform
\begin{equation}
 {V}_{\mu\nu;a}^{\Delta,\pm}(X^\mu;w,\bar w)=\frac{1}{2} \partial_{a} q_{\mu} \partial_{a} q_{\nu} \int_0^\infty d\omega \omega^{\Delta-1} e^{\pm i \omega q\cdot X-\varepsilon \omega}\,.
\end{equation}
Notice that $V_{\mu\nu}^{\Delta}$ does not transform covariantly under $SL(2,\mathbb{C})$ but the non-covariant terms are pure residual gauge, hence, following~\cite{Pasterski:2017kqt}, we will still call \eqref{VMellinSpin2} conformal primaries.

\subsubsection{Shadow Transform}
As in the spin-one case, we can perform a shadow transformation on the spin-two conformal primary wavefunction $h_{\mu \nu;a}^{\Delta}$~\eqref{hDelta} which takes its conformal dimension $\Delta$ to $2-\Delta$.
The expression for the shadow wavefunction was found in~\cite{Pasterski:2017kqt}:\footnote{For~\eqref{ShadowSpin2} to be well-defined at $-X^2=0$ and $q\cdot X=0$ one needs to prescribe a regulator analogous to the one in section~\ref{sec:softphoton}.}
\begin{equation}
\widetilde{h}_{\mu \nu;a}^{2-\Delta,\pm}= (-X^2)^{1-\Delta}h_{\mu \nu;a }^{2-\Delta,\pm}\,.
\label{ShadowSpin2}
\end{equation}
One can verify that the shadow wavefunction~\eqref{ShadowSpin2} satisfies the defining properties of spin-two conformal primary wavefunctions.

\subsubsection{Conformal Basis} 
A natural conserved inner product between complex spin-two wavefunctions is~\cite{Hawking:2016sgy,Ashtekar:1987tt,Crnkovic:1986ex,Lee:1990nz,Wald:1999wa}
\begin{equation}\label{IPSpin2}
( h_{\mu \nu}, h'_{\mu \nu})=-i\int d^3 X^i \Big[ h^{\mu \nu} \partial_0 h'^{*}_{\,\,\mu \nu}-2h^{\mu \nu} \partial_\mu h'^{*}_{\,\,0\nu}+h\partial^\mu h'^{*}_{\,\,0\mu}-h \partial_0 h'^{*}+h_{0\mu} \partial^\mu h'^{*} - (h \leftrightarrow h'^{*})\Big]\,,
\end{equation}
where $h=h^\rho_{\,\, \rho}$. It was shown in~\cite{Pasterski:2017kqt} that, with respect to the inner product \eqref{IPSpin2}, the gauge representative spin-two conformal primary wavefunctions ${V}_{\mu \nu;a}^{\Delta,\pm}$ form a complete $\delta$-function-normalizable basis on the principal continuous series $\Delta=1+i \lambda$ with $\lambda \in \mathbb R$:
\begin{eqnarray}\label{VVSpin2}
({V}_{\mu\nu;a}^{\Delta,\pm}(X^\mu;w,\bw),{V}_{\mu\nu;a'}^{\Delta',\pm}(X^\mu;w',\bw'))
=\pm (2\pi)^4 \,\delta(\lambda-\lambda')\delta_{aa'} \,\delta^{(2)}(w-w')\,,
\end{eqnarray}
up to zero-mode issues. Alternatively, an equally good basis of spin-two conformal primary wavefunctions is spanned by the shadow transformed conformal primaries $\widetilde{V}^\Delta$. As in the spin-one case, the primaries $V^\Delta$ are ill-defined at $\lambda=0$ and so is the inner product~\eqref{VVSpin2}. To obtain a complete basis of conformal primary wavefunctions on the principal continuous series $\Delta=1+i\lambda$ with $\lambda\in \mathbb{R}$ we need to include a canonically paired set of spin-two zero modes. One of them is the $\Delta \to 1$ limit of the conformal primary wavefunction $h^\Delta$ which becomes a pure diffeomorphism that can be recognized as the antipodally matched supertranslation of~\cite{Strominger:2013jfa,He:2014laa}. Hence $h^{\Delta=1}$ is the Goldstone mode associated to the spontaneously broken supertranslation symmetries. In the next section we will construct its canonical partner which in a momentum space decomposition is known to be the $\omega\to 0$ soft graviton~\cite{Strominger:2013jfa,He:2014laa}.

\subsection{\texorpdfstring{$\Delta=1$}{} Conformal Modes}
%\subsubsection{Goldstone Mode}
In the limit $\Delta\to 1$, the conformal primary wavefunction $h_{\mu \nu;a}^{\Delta}$~\eqref{hDelta} and its shadow $\widetilde{h}^{2-\Delta}_{\mu \nu;a}$~\eqref{ShadowSpin2} coincide and both reduce to a total derivative\footnote{Notice that there is another case where the primary wavefunction is pure gauge, namely for $\Delta=0$~\cite{Pasterski:2017kqt}.}.  We identify this pure diffeomorphism spin-two conformal primary with the Goldstone mode
\begin{equation}
h^{\rm G}_{\mu \nu;a}\equiv \lim_{\Delta \to 1}h_{\mu \nu;a}^{\Delta} = \partial_{\mu} \xi_{\nu; a}+\partial_{\nu} \xi_{\mu; a}\,,
\label{Goldstone1}
\end{equation}
where
\begin{equation}\label{xi1}
\xi_{\mu;a}\equiv-\frac{1}{8}\partial_{a}^2 [q_\mu \log(-q\cdot X)]\,.
\end{equation}
The diffeomorphism generator~\eqref{xi1} preserves the Lorenz gauge fixing of the conformal primary wavefunction and hence satisfies the harmonic gauge $\partial^\rho \partial_\rho \xi_{\mu;a}=0$.
Below it will be identified with a specific supertranslation. 

We seek a canonical partner for \eqref{Goldstone1}.  A new solution to Einstein's equations that is non-zero and not a total derivative is given by:
\begin{equation}
h_{\mu \nu;a}^{{\rm log},\pm} \equiv \lim_{\Delta \to 1}\partial_\Delta \left(h^{\Delta,\pm}_{\mu \nu;a}+\tilde{h}^{2-\Delta,\pm}_{\mu \nu;a}\right)  \,. 
\label{confsoftgraviton1}
\end{equation}
The next steps follow the ones developed in Section \ref{sec:csphoton}; we will not repeat the details here but directly gives the expression:
\begin{equation}\label{hST}
 h^{\rm ST}_{\mu \nu;a}=(q \cdot X) \log[X^2]h^{\rm G}_{\mu \nu;a} \delta(q\cdot X) +h^{\rm G}_{\mu \nu;a}\Theta\left(X^2\right)\,,
\end{equation}
which we will refer to as the supertranslation (ST) mode. It transforms as a $\Delta=1$  conformal primary wavefunction of weight (${3 \over 2},-{1 \over 2}$) for positive helicity. We will see below that it generates the inhomogeneous term in the supertranslations on $\scri^+$. 

\subsubsection{News Tensor on \texorpdfstring{$\scri^+$}{}}
We now want to express the $\Delta=1$ pure diffeomorphism~\eqref{Goldstone1} and the supertranslation mode~\eqref{hST} in retarded coordinates $(u,r,z,\bz)$ and expand them near $\mathcal I^+$.
The gravitational free data $C_{zz}(u,z,\bz)$ is given by the leading angular component of the metric
\begin{equation}
 h_{zz}(u,r,z,\bz)=r C_{zz}(u,z,\bz)+\dots\,,
\end{equation}
where the dots denote subleading terms in powers of $r$. The Bondi news tensor $N_{zz}$, which characterizes the outgoing gravitational radiation, is
\begin{equation}
 N_{zz} = \partial_u C_{zz}\,.
\label{news}
\end{equation}
The Bondi news is the gravitational analogue of the photon field strength $F_{uz}=\partial_u A_{z}$ at $\scri^+$.

For the $zz$ component of the pure diffeomorphism mode $h^{\rm G}_{\mu\nu;a}$ with $a=ww$ one finds the data\footnote{Notice that here the parametrization $q^\mu=\frac{1}{1+w \bw}(1+w\bw,w+\bw,-i(w-\bw),1-w\bw)$ for the null vector was taken.} 
\begin{equation}
\badat{2}\label{Czz}
C^{\rm G}_{zz;ww}=&\frac{(\bar z-\bar w)}{(z-w)^3(1+z \bz)(1+w \bw)}\,, \quad C_{\bz\bz;ww}^{{\rm G}}=\frac{\pi \delta^{(2)}(z-w)}{(1+z \bz)(1+w \bw)},\\
&\quad C^{\rm G}_{zz;ww}=-2D_z^2f\,, \quad C_{\bz\bz;ww}^{{\rm G}}=-2D_\bz^2f.
\eadat
\end{equation}
Here the function on the celestial sphere 
\begin{equation}
f=-\frac{(\bar z- \bar w)}{4(z-w)(1+z \bz)(1+w \bw)},
\end{equation}
is the supertranslation parameter. Near $\scri^+$, the vector field \eqref{xi1} becomes the supertranslation $f\partial_u$  and transforms as a vector of weight $({1 \over 2},-{1 \over 2})$. 
The Bondi news at $\scri^+$ for the supertranslation mode \eqref{hST} is
\begin{equation}
 N_{zz;ww}^{{\rm ST}}=-\frac{(\bz-\bw)\,\delta(u)}{(z-w)^3(1+z \bz)(1+w \bw)}\,, \quad N_{\bz\bz;ww}^{{\rm ST}}=-\frac{\pi \delta^{(2)}(z-w)\delta(u)}{(1+z \bz)(1+w \bw)},
\end{equation}
and similar expressions can be obtained for $a=\bw \bw$. The full spacetime associated with this asymptotic data is, at linear order, a gravitational analog of the electromagnetic shock wave geometry in Figure \ref{shock}. 

\subsubsection{Canonical Pairing}

We now show that the new $\Delta=1$ supertranslation mode is the canonical partner of the Goldstone mode with respect to the inner product~\eqref{IPSpin2}.
Taking the integral in~\eqref{IPSpin2} to be over $\scri^+$, we find that the inner product between the Goldstone and the $\Delta=1$ supertranslation mode  is:
\begin{equation}\label{symp1}
(h^{{\rm ST}}_{ww}(w), h^{\rm G}_{w'w'}(w'))_{\scri^+}=\frac{i\pi^2}{2}\gamma_{w\bw}\delta^{(2)}(w-w')\,,
\end{equation}
which implies that this mode generates the action of supertranslations on the Goldstone boson. 
To obtain the result above, we used that
\begin{equation}\label{int}
\int d^2 z  \frac{ ( w- z)(\bar w'-\bz)}{(\bar w-\bz)^3 (w'-z)^3}=\pi^2 \delta^{(2)}(w-w').
\end{equation}
The latter expression is found by noticing that the integral \eqref{int} takes the form of a conformal integral \cite{Dolan:2011dv}
\begin{equation}\label{In}
I_n=\frac{1}{2\pi} \int d^2 z \prod_{i=1}^n \frac{1}{(z-z_i)^{h_i}}\frac{1}{(\bz-\bz_i)^{\bar h_i}}\,,
\end{equation}
where $\sum_{i=1}^n h_i=\sum_{i=1}^n \bar h_i=2$, $h_i-\bar h_i \in \mathbb Z$.
Convergence of the integral requires $h_i +\bar h_i < 2$ for all $i$ although $I_n$ may be extended by analytic continuation. The integral~\eqref{In} was evaluated for $n=2$ in \cite{Dolan:2011dv}
\begin{equation}\label{I2}
I_2=\frac{\Gamma(1-h_1)\Gamma(1-h_2)}{\Gamma(\bar h_1)\Gamma(\bar h_2)}(-1)^{h_1-\bar h_1}2\pi \delta^{(2)}(z_1-z_2).
\end{equation}

From~\eqref{IPSpin2} it follows that the inner product of the $\Delta=1$ supertranslation mode and any other conformal primary $V^\Delta$ vanishes. Hence, the Goldstone mode and the conformal supertranslation mode,~\eqref{Goldstone1} and~\eqref{confsoftgraviton1}, are the pair of zero-modes that enhance the conformal primary wavefunctions $V^\Delta$ to a $\delta$-function-normalizable basis on the principal continuous series $\Delta=1+i \lambda$ with $\lambda \in \mathbb{R}$ including the point $\lambda=0$.

\subsection{Supertranslation Current}

The inner product allows us to associate to every classical solution of the linearized Einstein equations a mode of the quantum field operator $\hat{h}$. 
The quantum commutator of the operator conjugate to the Goldstone mode
\begin{equation}
 j_{ww}=i(\hat{h},h^{\rm G}_{ww}),
\end{equation}
is dimension $({3 \over 2},-{1\over 2})$ and generates the inhomogneous term for large diffeomorphisms on $\scri^+$:
\begin{equation}
 [j_{ww},\hat{h}_{zz}]=h^{\rm G}_{zz;ww}=2\partial_z \xi_{z;ww}\,.
\end{equation}
Using that $\hat C_{zz}=-2 D_z^2 \hat C$ and the expressions \eqref{Czz} for the Goldstone mode, we find
\begin{equation}
 j_{ww}=-2 \pi D_w^2 \Delta \hat{C}\,,
\end{equation}
where $\Delta \hat{C} \equiv \hat{C}_{\scri^+_+}-\hat{C}_{\scri^+_-}$.
We thus find that its derivative is related to (the $\scri^+$ part of) the supertranslation current of~\cite{Strominger:2013jfa,Barnich:2013axa}:
\begin{equation}
 P_z=4\pi D^z \hat C_{zz}|^{\scri^+_+}_{\scri^+_-} \,,
 \label{softop}
\end{equation}
via
\begin{equation}
4 D^w j_{ww} =P_w\,.
\end{equation}
$P_w$ is a spin-one operator of dimension $({3 \over 2}, {1\over 2})$ which nevertheless shares holomorphic properties of a conformally soft operator as follows. 
In an energy eigenbasis $P_w$ was shown to have the OPE with operators on the celestial sphere of energy $\omega$ 
\begin{equation} 
P_z{\cal O}_\omega(w)\sim {\omega \over z-w}{\cal O}_\omega(w).
\end{equation}
After Mellin transform to a conformal basis, this becomes 
\begin{equation} \label{dgl}
P_z{\cal O}_{(h, \bar h)}(w)\sim {1 \over z-w}{\cal O}_{(h+{1 \over 2}, \bar h+{1 \over 2})}(w).
\end{equation}
Hence, because OPEs with $P_z$ shift the conformal dimension of the operator by $({1 \over 2}, {1\over 2})$,  $P_z$ has Ward identities which are holomorphic in $z$ and  are equivalent to the leading soft graviton theorem. 

We expect the relation \eqref{dgl} may be a strong constraint on $S$-matrix elements in a conformal basis. 

\subsection{\texorpdfstring{$\Delta=2$}{} Goldstone Mode}
As noticed in~\cite{Pasterski:2017kqt}, the shadow conformal primary wavefunction \eqref{ShadowSpin2} with conformal dimension $2$ reduces to a total derivative. We identify this pure diffeomorphism spin-two conformal primary with the $\Delta=2$ ($\lambda=-i$) Goldstone mode:
\begin{equation}
\widetilde{h}^{\Delta=2}_{\mu \nu;a}\equiv -X^2 \, {h}^{2}_{\mu \nu;a } = \partial_{\mu} \zeta_{\nu; a}+\partial_{\nu} \zeta_{\mu; a}\,,
\label{Goldstone2}
\end{equation}
where ${h}^{2}_{\mu \nu;a }$ is the conformal primary \eqref{hDelta} with $\Delta=2$ and
\begin{equation}\label{xi2}
\zeta_{\mu; a}\equiv-\frac{1}{24} \partial_a^3 [X^\rho (q_\rho \partial_{\bar{a}} q_\mu- q_\mu \partial_{\bar{a}} q_\rho) {\rm log} (-q\cdot X)]\,.
\end{equation}
The diffeomorphism generator \eqref{xi2} preserves the Lorenz gauge fixing of the conformal primary wavefunction and hence satisfies the harmonic (de Donder) gauge $\partial^\rho \partial_\rho \zeta_{\mu;a}=0$.

We express the $\Delta=2$ pure diffeomorphism~\eqref{Goldstone2} in Bondi coordinates near $\mathcal I^+$ and find that its associated Bondi news tensor is given by
\begin{equation}\label{ShN2}
 \widetilde{N}^{\Delta=2}_{zz;ww}=\frac{1}{(z-w)^4}\,,
\end{equation}
and a similar expression can be obtained for $\widetilde{N}^{\Delta=2}_{\bz \bz;\bw \bw}$.
The news~\eqref{ShN2} is conformally soft as it transforms as a primary with conformal weights $(h,\bh)=(2,0)$ under an $SL(2,\mathbb C)$ transformation~\eqref{SL2Cspin2}.
Note that $\widetilde{N}^{\Delta=2}_{zz;ww}=D_z^3 Y^z_{w w}$ with
\begin{equation}\label{Yz}
Y^z_{ww}=-\frac{1}{6(z-w)}\,.
\end{equation}

In \cite{Kapec:2016jld}, a two-dimensional stress tensor for four-dimensional gravity was found whose $\mathcal I^+$ part is given by
\begin{equation}\label{Tzz}
 T_{ww}=2i\int du d^2 z \, \frac{\gamma^{z\bz}}{z-w}  u  D^3_z  \hat{N}_{\bz \bz}\,.
\end{equation}
Insertions of the operator \eqref{Tzz} into the tree-level S-matrix reproduce the Ward identity for a two-dimensional conformal field theory. The construction of this operator was recently generalized to $d>2$ in \cite{Kapec:2017gsg}.
After integration by parts, we find that the two-dimensional stress tensor \eqref{Tzz} is the convolution of the operator $\hat N$ and the $\Delta=2$ primary \eqref{ShN2}:
\begin{equation}
 T_{ww}=12i\int du d^2 z \, \frac{\gamma^{z\bz} }{(z-w)^4} \,u \hat{N}_{\bz \bz}=12i\int du d^2 z \gamma^{z\bz} \, \widetilde{N}^{\Delta=2}_{zz;ww} \,u \hat{N}_{\bz \bz}\,.
\end{equation}

\section*{Acknowledgements}
We are grateful to Thomas Dumitrescu, Gaston Giribet, Daniel Kapec, Blagoje Oblak, Sabrina Pasterski, Burkhard Schwab, Shu-Heng Shao and Alexander Zhiboedov for useful discussions. LD and AP acknowledge support from the Black Hole Initiative at Harvard University, which is funded by a grant from the John Templeton Foundation. LD was also supported by a Fellowship of the Belgian American Educational Foundation and by the CNRS and thanks the CPHT at Ecole Polytechnique for hospitality during her visits. This work was supported in part by DOE grant DE-SC0007870.

\appendix

\section{Inner Product}\label{app:IP}
We review here the result obtained in~\cite{Pasterski:2017kqt} that the spin-one conformal primary wavefunctions on the principal continuous series $\Delta=1+i \lambda$ with $\lambda \in \mathbb R$ are $\delta$-function-normalizable with respect to the inner product~\eqref{IPSpin1}.
Starting with the Klein-Gordon inner product between plane waves
\begin{equation}\label{KKplanewave}
 (e^{\pm i \omega q\cdot X}, e^{\pm i \omega' q'\cdot X})=\pm 2(2\pi)^3 \omega q^0 \, \delta^{(3)}(\omega q^i-\omega' q'^i)\,,
\end{equation}
the inner product for the representative~\eqref{VMellinSpin1} is obtained by taking two Mellin transforms:
\begin{equation}\label{KGVVspin1}
\badat{3}
&({V}_{\mu;a}^{\Delta,\pm}(X^\mu;w,\bar w),{V}_{\mu,a'}^{\Delta',\pm}(X^\mu;w',\bar w'))\\
&\qquad \qquad=\pm 2(2\pi)^3 \partial_a q \cdot (\partial_{a'} q')^* \int_0^\infty d\omega \,\omega^{i\lambda}\int_0^\infty d\omega'\, \omega'^{-i\lambda'} \omega \,q^0\delta^{(3)}(\omega q^i-\omega' q'^i)\\
 &\qquad \qquad=\pm (2\pi)^4  \delta(\lambda-\lambda') \delta_{aa'} \delta^{(2)}(w-w')\,,
\eadat
\end{equation}
where we used $q_0\delta^{(3)}(\omega q^i-\omega' q'^i)=\frac{1}{4 \omega^2}\delta(\omega-\omega')\delta^{(2)}(w-w')$, $\partial_a q \cdot (\partial_{a'}q)^*=2\delta_{aa'}$ and
\begin{equation}
\int_0^\infty d\omega \, \omega^{i\lambda-1}=2\pi \delta(\lambda).
\end{equation}
To compute the inner product for the spin-one conformal primary wavefunction $A^{\Delta,\pm}_{\mu;a}$~\eqref{prim1} note that the pure gauge mode can be expressed as the following Mellin transform
\begin{equation}
\Delta \partial_\mu \alpha_a^{\Delta,\pm} = (\partial_a q_\mu + q_\mu \partial_a) \frac{1}{(-q\cdot X\mp i \varepsilon)^\Delta}
= \frac{1}{(\mp i)^\Delta \Gamma(\Delta)}{V}_{\mu;a}^{\Delta,\pm}+ q_\mu \partial_a \int_0^\infty d\omega \, \omega^{\Delta-1} e^{\mp i\omega (-q\cdot X\mp i\varepsilon)} \,.
\end{equation}
This yields
\begin{eqnarray}
\badat{2}
(A_{\mu;a}^{\Delta,\pm}(X^\mu;w,\bw),A_{\mu;a'}^{\Delta',\pm}(X^\mu;w',\bw'))
&=\pm (2\pi)^4 \frac{\lambda \sinh(\pi \lambda)e^{\mp \pi \lambda}}{\pi(1+\lambda^2)}\,\delta(\lambda-\lambda') \delta_{aa'}\,\delta^{(2)}(w-w')\,.
\eadat
\end{eqnarray}
Notice that this inner product coincides with~\eqref{KGVVspin1} up to a normalization factor.

\section{Shadow Transform in the Embedding Space}~\label{app:shadow}
To compute the shadow wavefunction $\widetilde{A}_{\mu;a}^{2-\Delta}$ of the spin-one conformal primary wavefunction~\eqref{ADelta} it is convenient to use real coordinates $\vec{w}\in \mathbb{R}^2$ and use the embedding space formalism~\cite{Dolan:2011dv,SimmonsDuffin:2012uy}. The shadow $\widetilde{\mathcal{O}}_{a_1\dots a_{|J|}}(\vec{w})$ of the two-dimensional conformal primary $\mathcal{O}_{a_1\dots a_{|J|}}(\vec{w})$ in the symmetric
traceless rank-${|J|}$ representation of $SO(2)$ with dimension $\Delta$ is~\cite{Ferrara:1972xe,Ferrara:1972uq,Ferrara:1972ay,Ferrara:1973vz,Dolan:2011dv}
\begin{equation}
 \widetilde{\O}_{a_1\dots a_{|J|}}(\vec{w})=\frac{k_{\Delta,J}}{\pi} \int d^2 \vec{w}' \frac{1}{|\vec{w}-\vec{w}'|^{2(2-\Delta)}} \mathcal{I}_{a_1\dots a_{|J|},b_1,\dots b_{|J|}}(\vec{w}-\vec{w}')\mathcal{O}^{b_1 \dots b_{|J|}}(\vec{w}')\,,
 \label{SH2Dreal}
\end{equation}
where we take the normalization factor $k_{\Delta,J}=\Delta+J-1$ and $\mathcal{I}_{a_1\dots a_{|J|},b_1,\dots b_{|J|}}(\vec{w}-\vec{w}')$ is the inversion tensor for symmetric traceless tensors, formed from the symmetrised product of ${|J|}$ inversion tensors
\begin{equation}
 \mathcal{I}_{ab}(\vec{w}-\vec{w}')=\delta_{ab}-2\frac{(w_a-w'_a)(w_b-w'_b)}{|\vec{w}-\vec{w}'|^2}\,.
\end{equation}
The integral in~\eqref{SH2Dreal} is divergent unless $\Delta<1$ but can be extended to more general $\Delta$ by analytic continuation so that under conformal transformations~\eqref{SH2Dreal} defines a conformal primary operator in the symmetric traceless rank$-{|J|}$ representation of $SO(2)$ of weight $2-\Delta$~\cite{Dolan:2011dv}. 
The shadow operator $ \widetilde{\O}_{a_1\dots a_{|J|}}$ is most conveniently computed in terms of its uplift $ \widetilde{\O}_{\mu_1\dots \mu_{|J|}}$ to the embedding space $\mathbb{R}^{1,3}$ (recall that $q^\mu(\vec{w})\in \mathbb{R}^{1,3}$ and $-\frac{1}{2} q\cdot q'=|\vec{w}-\vec{w}'|^2$):
\begin{equation}
 \widetilde{\O}_{\mu_1\dots \mu_{|J|}}(\vec{w})=\frac{k_{\Delta,{J}}}{\pi} \int d^2 \vec{w}' \frac{\prod_{n=1}^{|J|}[\delta_{\mu_n}^{\nu_n} (-\frac{1}{2} q\cdot q')+\frac{1}{2} q'_{\mu_n} q^{\nu_n}]}{(-\frac{1}{2}q\cdot q')^{2-\Delta+{|J|}}}\O_{\nu_1\dots\nu_{|J|}}(\vec{w}')\,.
 \label{SH4D}
\end{equation}
The two-dimensional primary $\mathcal{O}_{a_1 \dots a_{|J|}}(\vec{w})$ is then recovered via the projection:
\begin{equation}
 \mathcal{O}_{a_1 \dots a_{|J|}}(\vec{w})=\frac{\partial q^{\mu_1}}{\partial w^{a_1}}\cdots \frac{\partial q^{\mu_{|J|}}}{\partial w^{a_{|J|}}} \mathcal{O}_{\mu_1 \dots \mu_{|J|}}(\vec{w})\,,
\end{equation}
and similarly for the shadow $\widetilde{\mathcal{O}}_{a_1 \dots a_{|J|}}(\vec{w})$.

The shadow transformed spin-one conformal primary $\widetilde{A}^{2-\Delta}_{\mu;a}$ is the projection of the uplifted shadow wavefunction $\widetilde{A}^{2-\Delta}_{\mu;\nu}$ computed from~\eqref{SH4D} by inserting the bulk-to-boundary propagator~\cite{Costa:2014kfa}
 \begin{equation}
A_{\mu;\nu}^\Delta(X^\mu;\vec{w})=\frac{(-q\cdot X)\eta_{\mu \nu}+q_\mu X_\nu}{(-q \cdot X)^{\Delta+1}}\,,
 \end{equation}
and using the identity
\begin{equation}
 \int d^2 \vec{w}' \frac{1}{|\vec{w}-\vec{w}'|^{2(2-\Delta)}} \frac{1}{(-q(\vec{w}')\cdot X)^\Delta}=\frac{\pi \Gamma(\Delta-1)}{\Gamma(\Delta)}\frac{(-X^2)^{1-\Delta}}{(-q(\vec{w})\cdot X)^{2-\Delta}}\,.
\end{equation}
This yields~\eqref{ShADelta} \cite{Pasterski:2017kqt}.
A similar computation yields the shadow transformed spin-two conformal primary~\eqref{ShadowSpin2}.

\section{Conformally Soft Modes in Bondi Coordinates}~\label{app:Bondi}
In this Appendix, we express the conformally soft modes in the retarded frame $(u,r,z,\bz)$ of Minkoswki spacetime $\mathbb{R}^{1,3}$. Cartesian coordinates $X^\mu$ with $\mu=0,1,2,3$ are related to Bondi coordinates $(u,r,z,\bz)$ by the transformation
\begin{equation}
 X^0=u+r\,, \quad X^1=r\frac{z+\bz}{1+z\bz}\,, \quad X^2=- i r \frac{z-\bz}{1+z\bz}\,, \quad X^3=r\frac{1-z\bz}{1+z\bz}\,,
\end{equation}
which maps the Minkowski line element to
\begin{equation}
ds^2=-du^2-2du dr+2r^2 \gamma_{z \bar z} dz d\bar z \quad \text{with}\quad \gamma_{z \bar z}=\frac{2}{(1+z \bar z)^2}\,.
\end{equation}
Lorentz transformations act on the Bondi coordinates as~\cite{Sachs:1962zza,Oblak:2015qia}
\begin{equation}
 u'=u \, K^{-1}(z,\bz)+\O(1/r)\,,\quad r'=r \, K(z,\bz)+\O(1)\,,\quad z'=\frac{a z+b}{c z+ d}+\O(1/r)\,,
\end{equation}
where
\begin{equation}
K(z,\bz)=\frac{|az+b|^2+|cz+d|^2}{1+z\bz}.
\end{equation}
The transformation of the $z$ coordinate expresses the fact that Lorentz transformations coincide with conformal transformations of the celestial sphere $\mathcal{C}\mathcal{S}^2$.

To give the explicit expressions of the conformal primaries we need the following expressions in the retarded frame:
\begin{eqnarray}
-X^2=u(2r+u)\,, \quad 
-q\cdot X=\frac{2r|z-w|^2}{(1+z \bz)}+u(1+w\bw)\,.
\label{X2qX}
\end{eqnarray}
The spin-one Goldstone mode for $a=w$ (positive helicity)
\begin{equation}
A^{\rm G}_{\mu;w}=\frac{\p_w q_\mu}{- q\cdot X}+\frac{(\p_w q \cdot X)q_\mu}{( q\cdot X)^2}
\end{equation} 
in the retarded frame is given by
\begin{eqnarray}
\badat{4}\label{Goldbulk}
A^{\rm G}_{u;w}(u,r,z,\bz)&=\frac{ -2 r(1+z \bw)(\bz-\bw)(1+z \bz)}{[2 r |z-w|^2 +u (1+z \bz)(1+w \bw)]^2}\,,\\
A^{\rm G}_{r;w}(u,r,z,\bz)&=\frac{2 u(1+z \bw)(\bz-\bw)(1+z \bz)}{[2 r |z-w|^2 +u (1+z \bz)(1+w \bw)]^2}\,,\\
A^{\rm G}_{z;w}(u,r,z,\bz)&=\frac{-2 r(2r+u) (\bz-\bw)^2}{[2 r |z-w|^2 +u  (1+z \bz)(1+w \bw)]^2}\,,\\
A^{\rm G}_{\bz;w}(u,r,z,\bz)&=\frac{2 r u (1+z \bw)^2 }{[2 r |z-w|^2 +u (1+z \bz)(1+w \bw)]^2}\,;
\eadat
\end{eqnarray}
similar expressions can be obtained for $a=\bar w$ (negative helicity). 
Its asymptotic behavior near null infinity $\scri^+$ gives components that fall off as $\mathcal O(1/r)$ or faster, except for the $z$ and $\bz$ components which have an $\mathcal O(1)$ piece at $\scri^+$. To make the $r-$expansion, we make use of the following formula (e.g. in \cite{Kutasov:1999xu}):
\begin{equation}\label{a2}
\left(\frac{y}{y^2+|z-w|^2}\right)^2\stackrel{y\to 0}{\approx} 2\pi \delta^{(2)}(z-w)+2\pi \,  y^2 \partial_z \partial_\bz\delta^{(2)}(z-w)+\frac{y^2}{|z-w|^4}+\mathcal O (y^{4}).
\end{equation}
with $y^2=\frac{u}{2r}(1+w \bar w)(1+z \bz)$, and we see that the denominator in \eqref{Goldbulk} expands in large $r$ as
\begin{equation}\label{deno}
\frac{1}{(2 r |z-w|^2 +u  (1+z \bz)(1+w \bw))^2}=\frac{\pi \delta^{(2)} (z-w) }{r u(1+z\bz)^2}+\frac{1}{4r^2|z-w|^4}+\frac{\pi \partial_z \partial_\bz\delta^{(2)}(z-w)}{2r^2}+\cdots
\end{equation}
This leads to
\begin{equation}
\badat{2}\label{GoldstoneScri}
 &A^{\rm G}_{z;w}=-\frac{1}{(z-w)^2}+\mathcal O(1/r)\,,\\
&A^{\rm G}_{\bz;w}=2\pi \delta^{(2)} (z-w)+\mathcal O(1/r).
\eadat
\end{equation}
The components of field strength \eqref{FCS} in retarded coordinates are found to be
\begin{equation}
\badat{2}
&F^{\rm CS}_{uz;w}=\frac{4r (r+u)(2r+u)(\bz-\bar w)^2 \delta(X^2)}{[2r |z-w|^2+u (1+w \bar w)(1+z \bz)]^2}+\frac{4r(r+u) (\bz-\bar w)^2 \delta(q \cdot X)}{u (1+z \bz)[2r |z-w|^2+u (1+w \bar w)(1+z \bz)]},\\[4pt]
&F^{\rm CS}_{u\bz;w}=\frac{-4r u (r+u)(1+z\bar w)^2\delta(X^2)}{[2r |z-w|^2+u (1+w \bar w)(1+z \bz)]^2}-\frac{4r(r+u) (1+z\bar w)^2 \delta( q\cdot X)}{(2r+u)  (1+z \bz)[2r |z-w|^2+u (1+w \bar w)(1+z \bz)]},\\[4pt]
&F^{\rm CS}_{ur;w}=\frac{-4 u (2r+u)(1+z\bar w)(1+z \bz)(\bz-\bar w)\,\delta(X^2)}{[2r |z-w|^2+u (1+w \bar w)(1+z \bz)]^2}-\frac{4(\bz-\bar w)(1+z\bar w)\,\delta(q\cdot X)}{[2r |z-w|^2+u (1+w \bar w)(1+z \bz)]},\\[4pt]
&F^{\rm CS}_{z\bz;w}=0,\\[4pt]
&F^{\rm CS}_{r z;w}=\frac{4ru(2r+u)(\bz-\bar w)^2 \delta(X^2)}{[2r |z-w|^2+u (1+w \bar w)(1+z \bz)]^2}+\frac{4r (\bz-\bar w)^2 \delta(q \cdot X)}{(1+z \bz)[2r |z-w|^2+u (1+w \bar w)(1+z \bz)]},\\[4pt]
&F^{\rm CS}_{r \bz;w}=\frac{-4ru^2(1+z\bar w)^2 \delta(X^2)}{[2r |z-w|^2+u (1+w \bar w)(1+z \bz)]^2}-\frac{4ru (1+z\bar w)^2 \delta(q \cdot X)}{(2r+u)(1+z \bz)[2r |z-w|^2+u (1+w \bar w)(1+z \bz)]}.\\[4pt]
\eadat
\end{equation}
Expanding the expressions above for large $r$ ($u$ fixed) and using that at $\scri$,
\begin{eqnarray}
\delta(X^2)=\frac{1}{2r}\delta(u)\, \virg 
\delta(q\cdot X)=\frac{(1+z \bz)}{2r}2\pi \delta^{(2)}(z-w)\Theta(-u),
\end{eqnarray}
with $\Theta(u>0)=1$ and $\Theta(u<0)=0$, we obtain the following values on $\scri$:
\begin{equation}
\badat{2}
&F^{\rm CS}_{uz;w}= \frac{\delta(u)}{(z-w)^2},\\
&F^{\rm CS }_{u \bz;w}= -2\pi \delta(u)\delta^{(2)}(z-w),\\
&r^2 F_{ur;w}^{{\rm CS}}=4\pi \gamma^{z\bar z}\Theta(-u)\partial_z \delta^{(2)}(z-w),\\
&F_{z\bz;w}^{{\rm CS}}= 0,
 \eadat
\end{equation}
where the $\mathcal O(1)$ piece of the $u\bz$ component comes from the leading piece in the expansion \eqref{deno} and where we used the identity $\delta^{(2)}(z)=-z\partial_z \delta^{(2)}(z)$ to obtain the third equation.

\bibliographystyle{style}
\bibliography{references}

\end{document}